\documentclass[fleqn,usenatbib]{mnras}

\usepackage{newtxtext,newtxmath}

\usepackage[T1]{fontenc}

\DeclareRobustCommand{\VAN}[3]{#2}
\let\VANthebibliography\thebibliography
\def\thebibliography{\DeclareRobustCommand{\VAN}[3]{##3}\VANthebibliography}

\usepackage{graphicx, subfigure}	
\usepackage{amsmath}	

\usepackage{scalerel,tikz}
\usetikzlibrary{svg.path}
\definecolor{orcidlogocol}{HTML}{A6CE39}
\tikzset{orcidlogo/.pic={
 \fill[orcidlogocol] svg{M256,128c0,70.7-57.3,128-128,128C57.3,256,0,198.7,0,128C0,57.3,57.3,0,128,0C198.7,0,256,57.3,256,128z};
 \fill[white] svg{M86.3,186.2H70.9V79.1h15.4v48.4V186.2z}
 svg{M108.9,79.1h41.6c39.6,0,57,28.3,57,53.6c0,27.5-21.5,53.6-56.8,53.6h-41.8V79.1z M124.3,172.4h24.5c34.9,0,42.9-26.5,42.9-39.7c0-21.5-13.7-39.7-43.7-39.7h-23.7V172.4z}
 svg{M88.7,56.8c0,5.5-4.5,10.1-10.1,10.1c-5.6,0-10.1-4.6-10.1-10.1c0-5.6,4.5-10.1,10.1-10.1C84.2,46.7,88.7,51.3,88.7,56.8z};
}}
\newcommand\orcidicon[1]{\href{https://orcid.org/#1}{\mbox{\scalerel*{
\begin{tikzpicture}[yscale=-1,transform shape]
\pic{orcidlogo};
\end{tikzpicture}
}{|}}}}

\title[Photometric Binaries]{Photometric Determination of Main-Sequence Binaries with Gaia}

\author[A.~L.~Wallace et al.]{A.~L.~ Wallace$^{\orcidicon{0000-0002-6591-5290}\,1}$\thanks{E-mail: alex.wallace1@monash.edu}\\
$^{1}$School of Physics \& Astronomy, Monash University, Victoria, Australia\\}

\pubyear{2023}

\begin{document}
\label{firstpage}
\pagerange{\pageref{firstpage}--\pageref{lastpage}}
\maketitle

\begin{abstract}
Since its launch in 2013, the Gaia space telescope has provided precise measurements of the positions and magnitudes of over 1 billion stars.  This has enabled extensive searches for stellar and sub-stellar companions through astrometric and radial velocity measurements.  However, these surveys require a prior knowledge of any unresolved companion affecting the results which can be identified using photometry.  In this work, Gaia's magnitude measurements are combined with near-infrared observations from 2MASS and WISE and simulation-based inference is applied to constrain astrophysical parameters and search for hidden companions.  This method is first tested on simulated sets of binary stars before expanding to Gaia's non-single star catalogue.  Using this test, a region is identified on the H-R diagram in which the method is the most accurate and all Gaia sources within that region are analysed.  This analysis reproduces a known anti-correlation between metallicity and binary fraction.  Finally, the method is applied to the nearby star cluster M67 and, using previous studies of the metallicity distribution, it is possible to improve constraints on binary fraction.  From this the binary fraction in the cluster is calculated to vary from 30\% in the outer cluster to 45\% near the core.  This is found to be significantly higher the 23\% binary fraction calculated for the wider stellar neighbourhood.
\end{abstract}




\section{Introduction}
\begin{figure}
    \centering
    \includegraphics[width=1.0\linewidth]{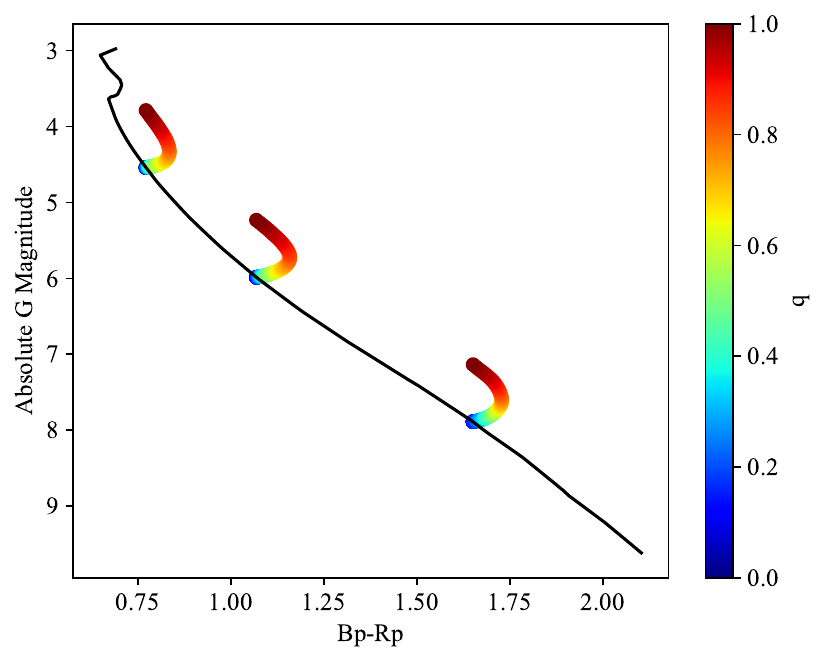}
    \caption{Positions of binaries on the H-R diagram for primary masses of 0.6, 0.8 and 1.0\,M${\odot}$.  The isochrone for single main-sequence stars of solar age (4.6\,Gyr) and metallicity ([Fe/H]=0) is also shown.}
    \label{fig:binary_hr}
\end{figure}

It has long been understood that a significant fraction of stars in our galaxy form in multiple systems.  Previous studies show about 10\% of field stars are hosting at least one companion, a fraction which more than doubles in star clusters and star-forming regions \citep{Duchene18}. This substantial difference in binary fraction between different stellar populations is of particular interest when studying stellar evolution.  By studying stellar multiplicity in a wide range of environments such as star clusters and the Galactic field, and taking dynamical interactions into account, this can help test theories of star formation \citep{Duchne_2013}.  For instance, it has been established that wide binaries (>100\,AU) are rare in young star clusters \citep{deacon2020} leading to different theories of stellar interactions and evolution to explain this discrepancy.\\
Studies of tight binary systems (<100\,AU) are important for our understanding both stellar evolution and planet formation \citep{saleh2009}.  It is understood that tight binary fraction, like total binary fraction, increases with the mass of the primary star \citep{clark2012} and that the chemical properties of a star may play an even larger role in determining stellar multiplicity \citep{badenes2018,mazzola2020}.  Due to the many factors such as a star's mass, composition and neighbourhood conditions affecting the binary fraction, it is essential to form a highly accurate picture of binary fraction as a function of these properties.  Studies of tight binaries has only been made possible in recent decades by advances in spectroscopy and radial velocity measurements.  However, with highly accurate photometry and astrometry, it is also possible to determine stellar multiplicity from a star's observed colour and magnitude.\\
Since the launch of the Gaia Space Telescope, we now have precise measurements of the positions of over 1 billion stars providing a detailed 3-dimensional map of the Milky Way.  The third data release (DR3), based on the first 34\,months of observations \citep{gaia_review}, contains a wealth of new information on existing sources including precise radial velocity (RV) and astrometric observations.  These observations are useful in searching for stellar companions and have successfully identified previously unknown multiple star systems \citep{mugrauer2022}.\\
As well as RV and astrometry, Gaia has also provided us with highly accurate magnitudes and colours of stars in the Milky Way.  Using these magnitudes, it is possible to identify unresolved stellar-mass companions to main-sequence stars.  Stellar multiplicity on the main-sequence can be observed on the H-R diagram, as shown in Figure~\ref{fig:binary_hr}, for three different primary masses.  The isochrone for a single atar of solar age and metallicity is shown for comparison.

As shown in Figure~\ref{fig:binary_hr}, as the binary mass ratio increases, the star becomes redder and slightly brighter.  Above a mass ratio of $\sim$0.7, the star's colour shifts to shorter wavelengths and the brightness increases rapidly, with an equal binary having a magnitude $\sim$0.75 brighter than a single star.\\
Unresolved stellar companions are an important factor to consider when conducting astrometric surveys with Gaia as the presence of an additional high-mass companion could hinder the search for lower mass companions if not taken into account.  By independently investigating binarity early on, we can remove these potential obstacles from future astrometric surveys.  However, a major challenge in identifying binaries from photometry alone is the multiple astrophysical parameters, such as metallicity, primary mass and age, all affecting a star's position on the H-R diagram.  Previous studies have constrained stellar properties of binaries \citep{Traven_2020} but these were binaries that had already been identified through other methods.  Other studies have investigated how photometry can determine binarity from photometry \citep{hartigan2001} but did not constrain other stellar parameters.\\
In this study, simulation-based inference is applied to photometric measurements from Gaia to constrain mass ratios of potential binaries as well as the primary mass, metallicity and age.  Degeneracies are minimised by also including near infrared photometry from the 2MASS and WISE surveys.  For the purposes of this study, the sample is restricted to the main sequence due to higher numbers of potential targets and more constrained stellar models for this evolutionary stage.  In the following sections, the method is introduced and first tested on simulated data before being applied to a selection of nearby star clusters and a large sample across the entire sky from the Gaia DR3 catalogue to paint a clearer picture of stellar multiplicity in the local neighbourhood.
\section{Method}
\label{sec:method}
This section describes the method in which astrophysical parameters (primary mass, mass ratio, metallicity, age and distance) are calculated based on observed magnitudes.  As multiple parameters are recovered from multiple magnitudes, astrophysical parameters are denoted by the vector
\begin{equation}
\vec{\theta}=\left[M_{1},q,\tau,\left[\mathrm{Fe/H}\right],d\right]\label{eq:theta}
\end{equation}
where $M_{1}$ is the primary mass, $q$ is the mass ratio (always between 0 and 1), $\tau$ is the age of the system in Gyr, [Fe/H] is the metallicity relative to the Sun and $d$ is the distance in pc.  We denote the observed quantities with the vector
\begin{equation}
\begin{aligned}
    \vec{x}=[ m_{B},m_{G},m_{R},m_{J},m_{H},m_{K},m_{W1},m_{W2},\varpi,\\
    \sigma_{B},\sigma_{G},\sigma_{R},\sigma_{J},\sigma_{H},\sigma_{K},\sigma_{W1},\sigma_{W2},\sigma_{\varpi}]
\end{aligned}
\label{eq:x}
\end{equation}
where the $m$'s are the apparent magnitudes in the Bp, G, Rp, J, H, Ks, W$_{1}$ and W$_{2}$ wavelength bands, $\varpi$ is the parallax in mas measured by Gaia and the $\sigma$'s are the associated errors in these observations.\\

In order to recover astrophysical parameters $\vec{\theta}$ from observations $\vec{x}$, there are several approaches we could take, which rely on possessing an accurate model of $\vec{x}$ as a function of $\vec{\theta}$.  The simplest method would be to construct a likelihood function which models the probability a certain set of parameters produce the observed magnitudes.  A set of parameters which produces the maximum value of this probability would then be taken as the true parameters.  However, the relation between $\vec{\theta}$ and $\vec{x}$ is highly degenerate (e.g. a metal-rich single star could have similar magnitudes to a metal-poor binary) which produces local maxima and minima in the likelihood function.  When scanning the likelihood function there is then a risk of getting stuck on a local maximum with parameters far away from the true $\vec{\theta}$.  In order for this method to work, an initial guess for $\vec{\theta}$ is required to be implausibly close to the true value.\\

The aim of this study is to obtain the posterior distributions of the astrophysical parameters $p(\vec{\theta}|\vec{x})$ which is a measure of the probability of a set of parameters $\vec{\theta}$ for given data $\vec{x}$.  This distribution is generally expressed by Bayes' Theorem:
\begin{equation}
    p(\vec{\theta}|\vec{x}) = p(\vec{x}|\vec{\theta})p(\vec{theta})
\end{equation}
where $p(\vec{x}|\vec{\theta})$ is the probability of the data $\vec{x}$ for given parameters $\vec{\theta}$ (the likelihood) and $p(\vec{theta})$ is the prior or expected distribution of the parameters.  This study relies on simulation-based inference (SBI) \citep{cranmer2020}, also known as likelihood-free inference which aims to approximate the posterior distribution with no assumptions about the form of the likelihood function.  SBI works by using a set of parameters $\vec{\theta}$ to generate mock data $\vec{x}$ and training a neural network to determine a relation between $\vec{\theta}$ and $\vec{x}$.  This relation is then used to calculate an $N$-dimensional probability function for a given set of observations, where $N$ is the number of parameters.  This method is faster than a likelihood maximisation as we rely on an overall distribution rather than a point-by-point comparison and, as we can estimate the density in parameter space, we can concentrate on the most likely values for $\vec{\theta}$.
\subsection{Simulated Training Set}
Before analysing real data, a simulated set of stars is constructed with given $M_{1}$, $q$, $\tau$, [Fe/H] and $d$.  This is the set of astrophysical parameters denoted $\vec{\theta}$ (Equation~\ref{eq:theta}).  For this `training' set, 100,000 systems are simulated, with M$_{1}$ ranging from 0.4--5\,M$_{\odot}$, $q$ from 0--1, $\tau$ from 0.2--10\,Gyr and [Fe/H] from -2 to +0.5.\\

The values for $M_{1}$ and [Fe/H] in the training set were sampled from beta distributions designed to approximate the initial mass function and an expected metallicity distribution which is maximised near solar metallicity.  This distribution has a probability density function given by:
\begin{equation}
    p(x;\alpha,\beta)\propto x^{\alpha-1}(1-x)^{\beta-1}
\end{equation}
where x is the parameter (either $M_{1}$ or [Fe/H]).  The mass distribution has $\alpha=1$, $\beta=5$ and the metallicity distribution has $\alpha=10$, $\beta=2$.  The values of $q$ and $\tau$ are sampled from uniform distributions.\\

\begin{figure}
    \centering
    \includegraphics[width=1.0\linewidth]{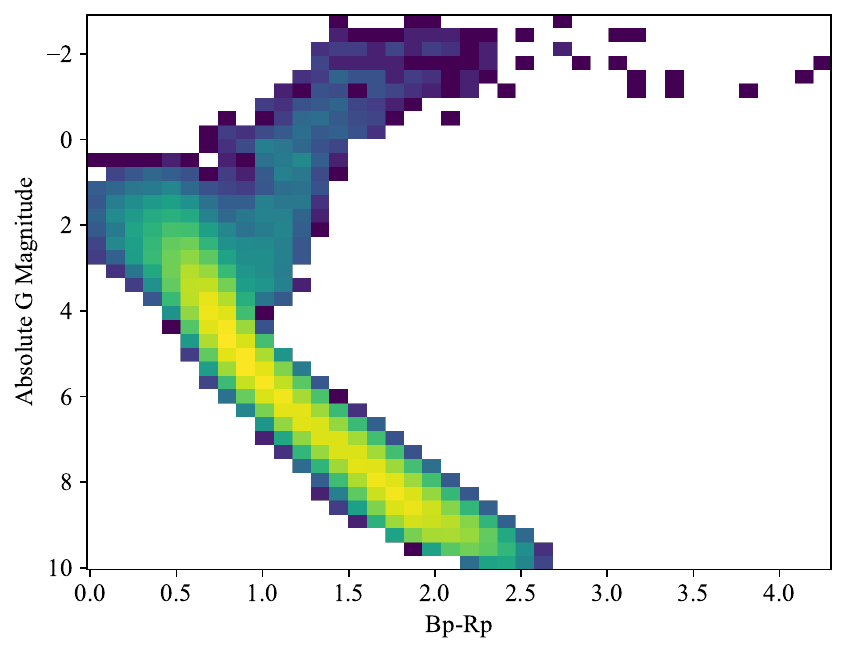}
    \caption{Histogram of the simulated training set on the H-R diagram before the addition of errors.}
    \label{fig:sim_hist}
\end{figure}

These stars are simulated with the \texttt{isochrones} package \citep{morton2015} to convert the astrophysical parameters $\vec{\theta}$ into a set of absolute magnitudes in the Gaia (Bp,G,Rp), 2MASS (J,H,Ks) and WISE (W$_{1}$,W$_{2}$) bands.  If a star produces an invalid result (e.g. it is too old for its mass) it is discarded and new stars are simulated until 100,000 valid results are produced.  This trainingn set initially ignores distance and only simulates absolute magnitudes in all bands.  The training set of 100,000 stars are shown as a histrogram on the H-R diagram in Figure~\ref{fig:sim_hist}.\\
After producing the sample set of absolute magnitudes.  Each star is assigned 100 distances between 1--1000\,pc sampled from a uniform distribution.  The absolute magnitudes are then converted to apparent magnitudes.  The training now consists of 10 million stars, each with a parallax and apparent magnitudes in eight bands.
\subsubsection{Uncertainties}
\begin{figure*}
    \centering
    \includegraphics[width=0.8\linewidth]{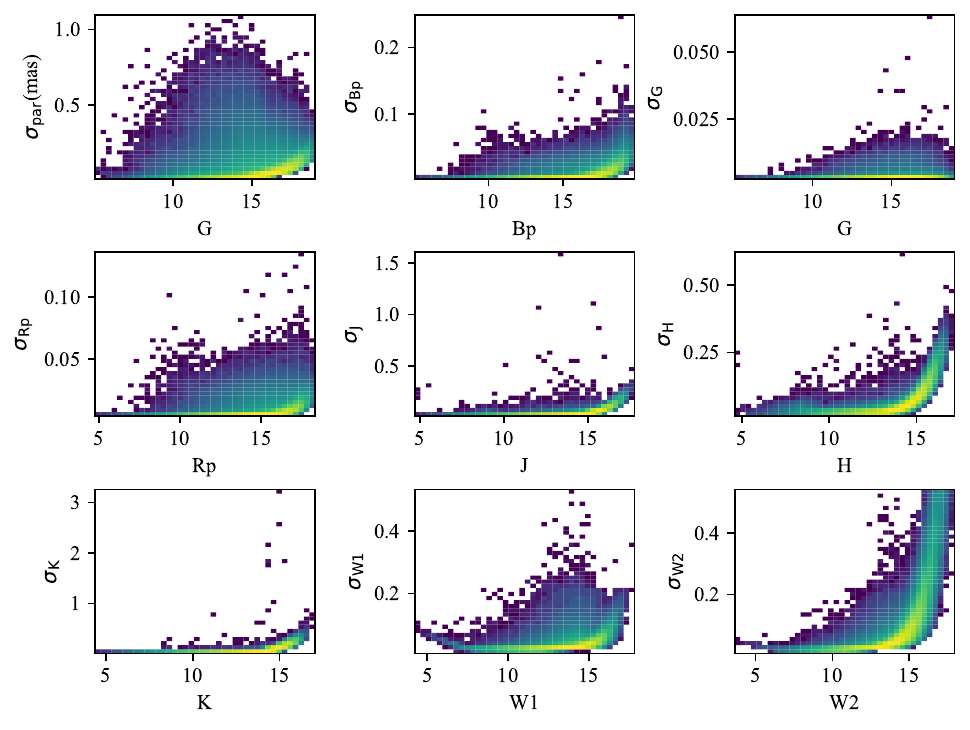}
    \caption{Distributions of magnitudes and associated errors, shown on a log density scale from dark blue (lowest) to yellow (highest).  Note parallax error is shown as a function of G magnitude rather than parallax as the magnitude is a key factor determining astrometric errors (and therefore errors in parallax)} \citep{lindegren21}.
    \label{fig:err_real}
\end{figure*}

Uncertainties in parallax and magnitude are handled using a similar method to \citet{Hahn_2022}, in which uncertainties are assigned to the simulated data and treat these as extra variables when running the SBI.  These uncertainties are calculated using real observed uncertainties from Gaia, 2MASS and WISE.  These observations give an indication of how the uncertainties should vary as a function of magnitude.  The distributions for the real data are shown in Figure~\ref{fig:err_real}.
Using this real data, a grid of colours (Bp-Rp) and apparent G magnitude is constructed.  Each simulated source is sorted into a square on this grid based on its simulated colour and magnitude.  A real source is randomly selected from this square and its errors in parallax and magnitude are applied to the simulated star.  This ensures that the simulated source will have a set of errors that is consistent with real data, rather than a set of independently generate random errors.\\

In total, this simulation produces 18 `observables' for each star (as shown in Equation~\ref{eq:x}): a parallax, apparent magnitudes in Gaia bands ($Bp,G,Rp$), 2MASS bands ($J,H,Ks$) and WISE bands ($W_{1},W_{2}$) and their associated uncertainties.  The set of observables is denoted $\vec{x}$.
\subsection{Parameter Inference}
\begin{figure}
    \centering
    \includegraphics[width=0.8\linewidth]{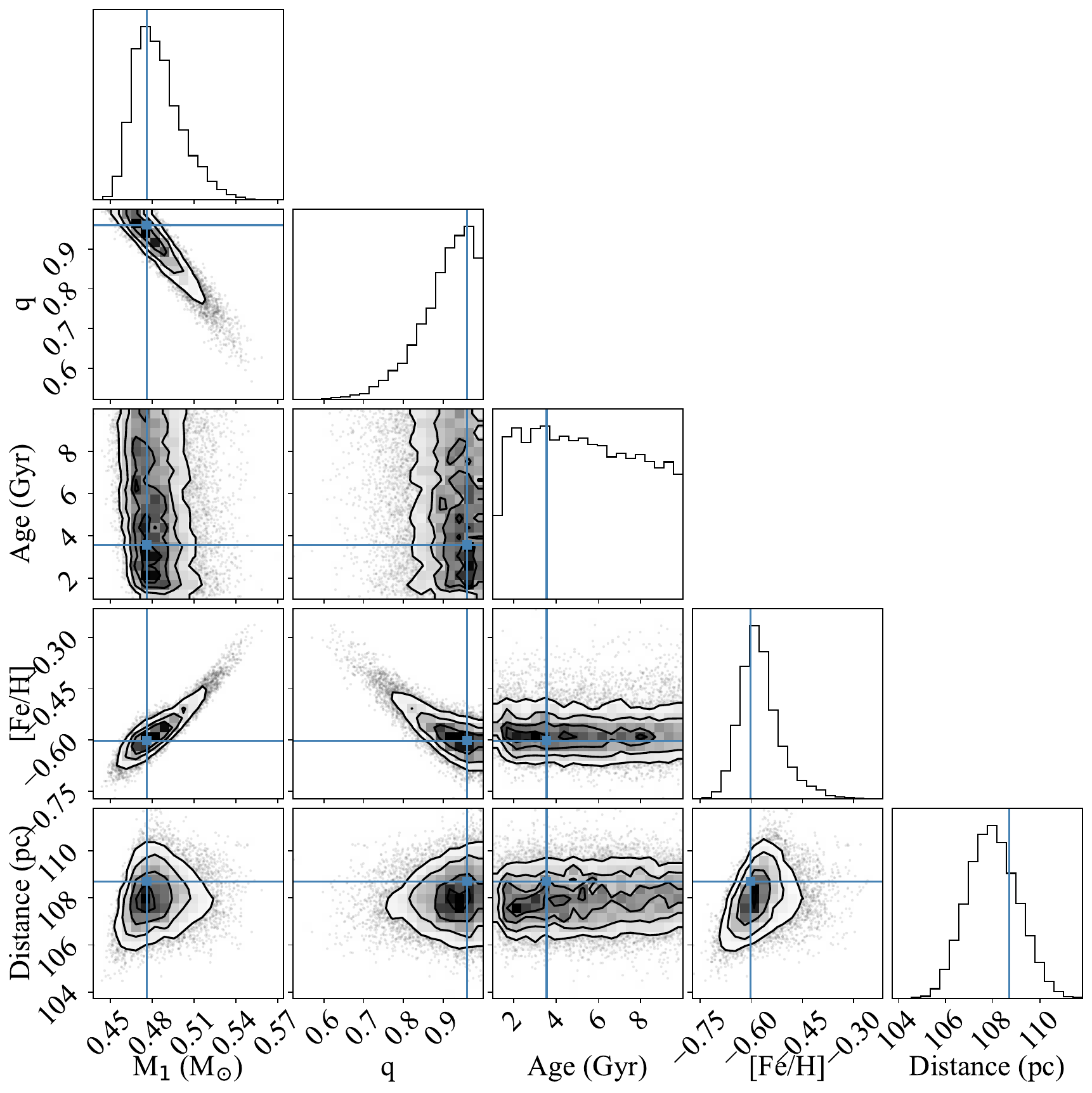}
    \caption{Example corner plot showing parameter distributions for a simulated binary star and correlation between parameters.  In this example, a set of observations are calculated using \texttt{isochrones} for a set of given astrophysical parameters.  The SBI method is then applied to these observations in an attempt to reproduce the simulated parameters.  The simulated `true' values are then shown for comparison.}
    \label{fig:pairplot_ex}
\end{figure}

The set of astrophysical parameters $\vec{\theta}$ which map to observables $\vec{x}$ are then used to train a neural network which is used to derive the full parameter space and approximate the posterior distribution $p(\vec{\theta}|\vec{x})$ where $\vec{x}$ is a set of observations, either real or simulated.\\
An example corner plot of the posteriors for a simulated binary is shown in Figure~\ref{fig:pairplot_ex}.
As shown in Figure~\ref{fig:pairplot_ex}, the SBI method in this particular case has succeeded in accurately recovering the correct masses, metallicity and distance while the star's age has proven difficult to constrain as its colour and magnitude change little during the main sequence phase.\\
The models outlined above don't take extinction into account and the following tests on simulated data assume no extinction.  However, when applying the method on real data, extinction can have a significant effect on a star's colour and magnitude such that, if left unchecked, parameter inference will be highly inaccurate.  To overcome this, extinction curves for each band are applied.  These are calculated according to a polynomial function of the star's colour and extinction at 550\,nm which is provided for each star in the Gaia catalogue using models from \citet{fitz99}.  After calculating the extinction in each band, this is simply subtracted from the measured magnitude.  This new magnitude is then used in SBI to recover astrophysical parameters and extinction can be safely ignored.
\section{Testing of the Model}
\subsection{Initial Test with Simulated Data}
\label{sec:init_test}
\begin{figure}
    \centering
    \includegraphics[width=0.8\linewidth]{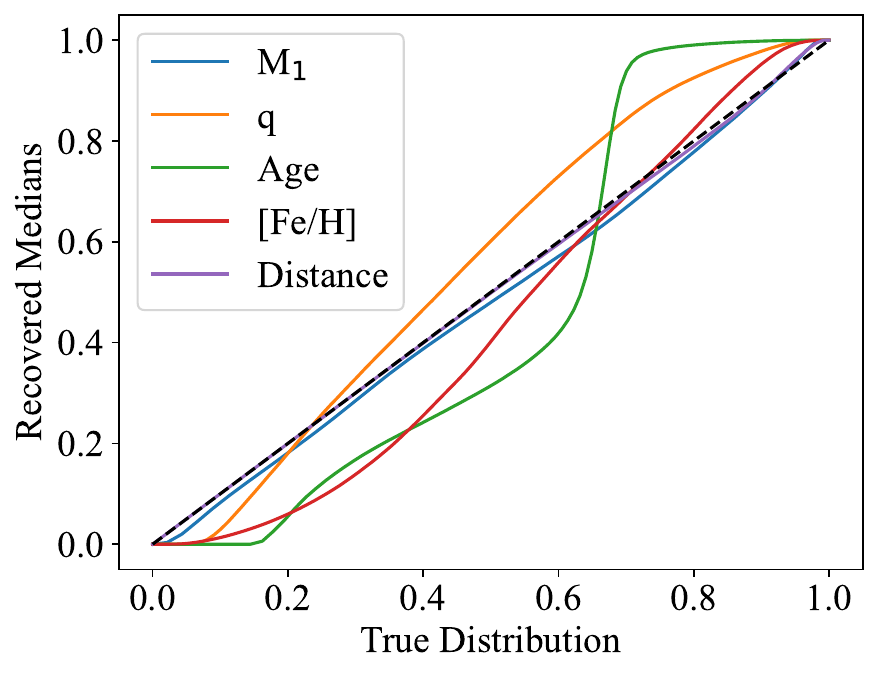}
    \caption{P-P plot of the cumulative distributions of the recovered values against true values.  Dashed line is 1-1 line where the distributions are the same.}
    \label{fig:ppplot}
\end{figure}

\begin{figure*}
    \centering
    \includegraphics[width=0.8\linewidth]{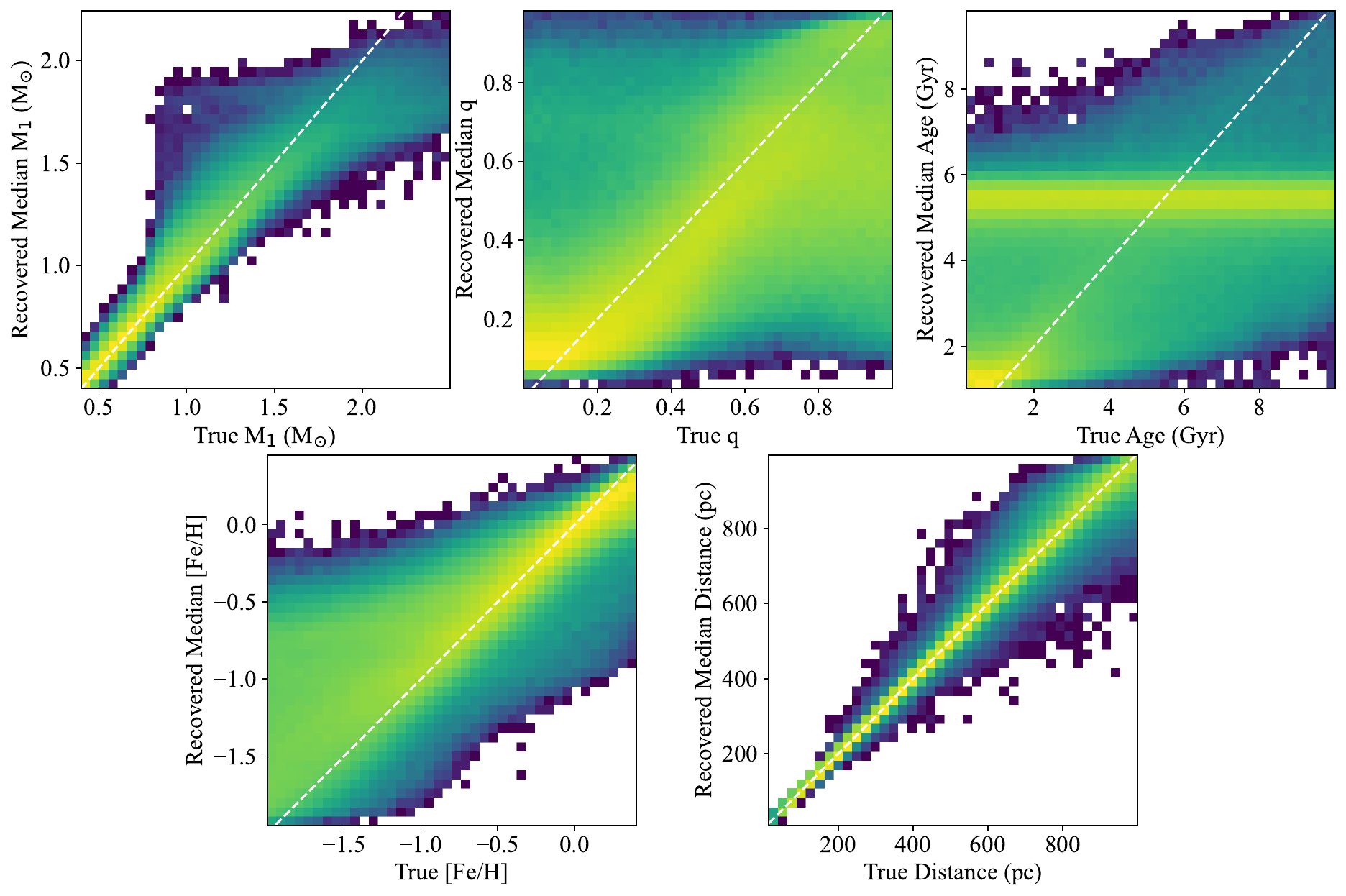}
    \caption{Comparison histograms between the simulated values of the parameters and recovered medians.  Density is on a log scale.  The dashed line represents 1-1 mapping between simulated and recovered values.}
    \label{fig:compare_all}
\end{figure*}

In order to test how this method is performing overall, SBI was run on a sample of 1\,million simulated stars with all parameters sampled from uniform distributions.  Each source was sampled 2,000 times and the median values of these samples were compared to the simulated `true' values.  One way of comparing the simulated parameter distributions is with a probability-probability (P-P) plot.  This plots the normalised cumulative distribution of each recovered parameter against the true cumulative distribution \citep{gibbons2014nonparametric}.  If the distributions agree perfectly, this plot follows a straight 1-1 line.  A P-P plot for the primary mass, mass ratio, age, metallicity and distance distributions is shown inFigure~\ref{fig:ppplot}.
The P-P plot demonstrates that the recovered distributions of primary mass and distance correspond almost exactly with the simulated distribution.  By contrast, the recovered age distribution is very different as this parameter is the most difficult to constrain by photometric methods.  The P-P plot for the mass ratio is almost always above the 1-1 plot which indicates this parameter is frequently underestimated.  The metallicity plot is mostly below the 1-1 plot indicating this parameter is usually overestimated.  This could be due to the degeneracies when mapping parameters to observed magnitudes.  There could, therefore, be low metallicity binaries which are incorrectly labelled as high metallicity single stars.  The distributions of each parameter are shown in detail in Figure~\ref{fig:compare_all} which plots 2D histrograms of recovered medians against simulated values.
As shown in Figure~\ref{fig:compare_all}, similar to the P-P plot, strong correlation can be seen between true and recovered values of primary mass and distance.  Most of the recovered mass ratio values are close to the true values but more than half of them have been underestimated.  Most of the recovered values of metallicity agree with the simulated values but there are many for which the metallicity has been overestimated.  As stated above, this is likely due to degeneracies and could also be due to the fact that the training set is denser at higher metallicities leading the SBI recovery to favour stars at these higher metallicities.\\
\subsection{Recovered Binary Fractions}
\label{sec:binary_recovery}
\begin{figure}
    \centering
     \subfigure[Binary fraction for stars with simulated $q<0.5$]{\label{fig:false_binary} \includegraphics[width=1\linewidth]{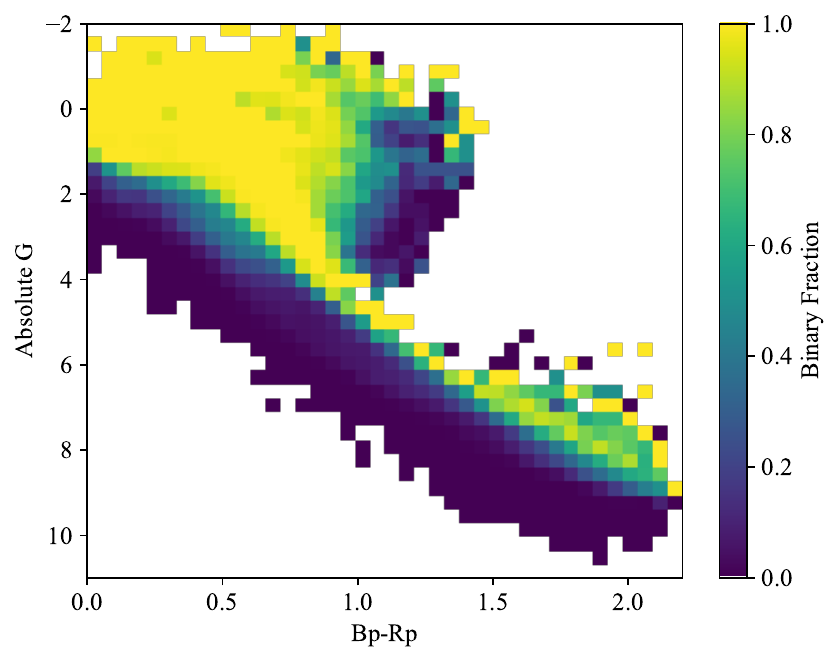}}
    \subfigure[Binary fraction for stars with simulated $q>0.5$]{\label{fig:true_binary} \includegraphics[width=1\linewidth]{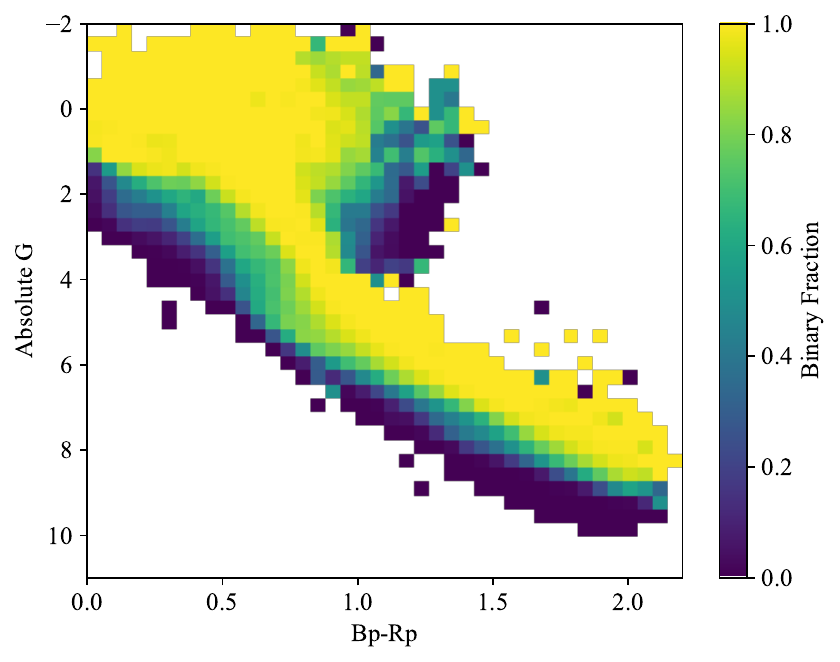}}
    \caption{Fractions of simulated single stars and binary stars recovered as binaries shown on the H-R diagram.}
    \label{fig:bin_both_hr}
\end{figure}

Of particular scientific interest is the prevalence of binary systems in stellar populations.  For the purposes of this study, a `binary' is defined as a source with $q \geq 0.5$.  Below this mass ratio, it is unlikely a star's binarity can be determined by photometry as the magnitudes are not sufficiently brighter than the single star case (as indicated in Figure~\ref{fig:binary_hr}.)  A `true' binary is defined as a simulated source with true $q \geq 0.5$ and a `recovered' binary as a source with recovered median $q \geq 0.5$. 
Using this definition of a binary, the relationship between recovered binary fraction and colour and magnitude was investigated.  From this, it is possible to identify the regions of the H-R diagram at which this method is most sensitive and accurate as well as those regions in which single stars are incorrectly recovered as binaries.  The binary fractions for the sample of true singles (true $q<0.5$) and true binaries are shown on H-R diagrams in Figure~\ref{fig:bin_both_hr}.

Figure~\ref{fig:true_binary} shows a clear `stripe' along the main sequence along which binaries are successfully recovered.  This demonstrates the limitation of photometry as, below this stripe, the secondary star is not bright enough to significantly affect the colour and magnitude.  The stripe also gets less pronounced at the lower mass end of the main sequence, indicating fewer binaries are recovered for primary stars at these masses.  This is probably due to companions not significantly affecting the magnitude until $q$ is very high (>0.8) and the larger magnitude uncertainties associated with fainter stars.\\
It should be noted that the binary fraction diagram for true singles shown in Figure~\ref{fig:false_binary} also exhibits a stripe along the main sequence, however it is confined to brighter magnitudes and is less pronounced.  This stripe is due to degeneracies in the relation between parameters and magnitudes.  Many of the single stars falsely recovered as binaries have high mass and/or metallicity which, when an uncertainty is added, can make the star `look' like a binary.\\
Both H-R diagrams show a region at the top end of the main sequence (and where stars are moving towards the giant branch) in which almost 100\% of stars are recovered as binaries.  In this region, there is very little difference between Figures~\ref{fig:false_binary} and~\ref{fig:true_binary}.  This is an example of where the method fails, which is most likely caused by the relative sparsity of sources in the training set as shown in Figure~\ref{fig:sim_hist}.\\
\subsection{Non-Single Star Catalogue}
\begin{figure}
    \centering
    \includegraphics[width=1.0\linewidth]{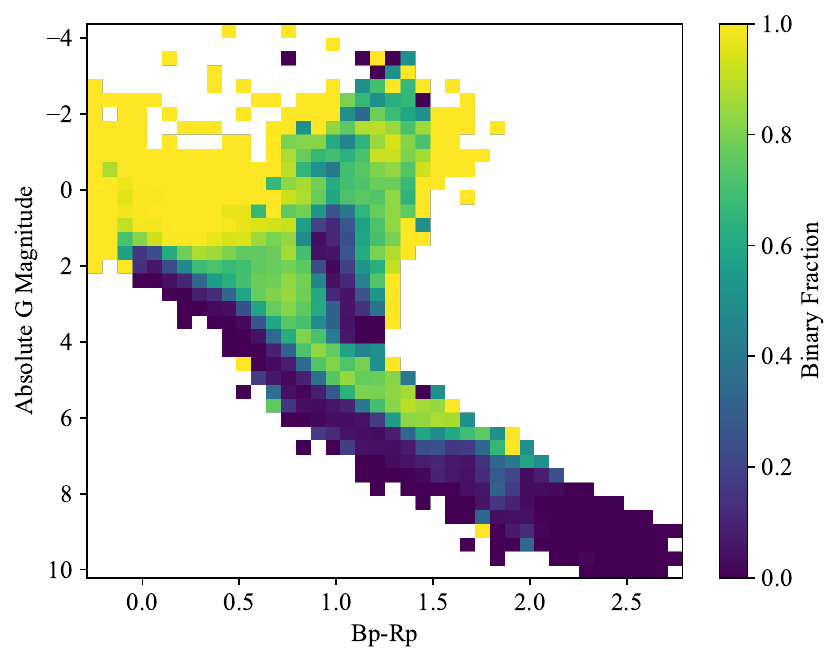}
    \caption{Binary fraction on an H-R diagram for Gaia's NSS catalogue.}
    \label{fig:bf_nss}
\end{figure}

In order to further test the sensitivity of the recovery of binary fraction, the SBI method was also run on the non-single star tables (NSS) from Gaia DR3.  These are stars which have been identified as astrometric, spectroscopic and eclipsing binaries.  For each star in the catalogue, the SBI was run as previously described and calculated which sources had median recovered $q\geq 0.5$.  The binary fraction on an H-R diagram is shown in Figure~\ref{fig:bf_nss}.
The result from the non-single star catalogue reveals a region of the H-R diagram where binaries differ from singles so little in their photometry that they are undetectable.  Most of these would have low mass ratios or are at the fainter end of the main sequence where magnitude uncertainties are larger.  All results from Figures~\ref{fig:bin_both_hr} and~\ref{fig:bf_nss} show a region at the brighter end of the main sequence where all stars are recovered as binaries.  From these results, a region of the H-R diagram can then be identified in which the method is most effective.  When running the method on a large sample of Gaia data, this study can then be restricted to this region in which the method sensitive to binaries with minimal contamination from `false' binaries.
\section{Large Gaia Sample}
\begin{figure}
    \centering
    \includegraphics[width=1.0\linewidth]{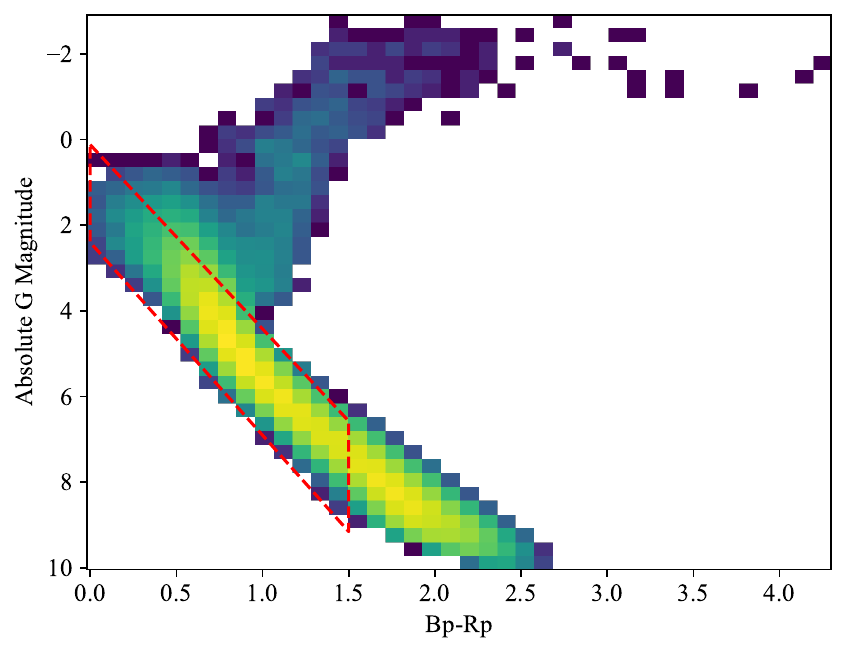}
    \caption{Region of the H-R diagram in which this study is focused, shown as a red-dashed box on the distribution of the training set shown in Figure~\ref{fig:sim_hist}.}
    \label{fig:box}
\end{figure}

When running SBI on a large sample of Gaia data, the sample is restricted to a stripe along the main sequence in which this method has shown to be effective at recovering binary fractions.  This region is defined by a parallelogram on the H-R diagram restricted to stars with Bp-Rp between 0--1.5 with absolute G magnitude between two lines 2.2 magnitudes apart with a slope of 4.5.  The region studied is shown in Figure~\ref{fig:box}.
Additionally, conditions are set on the magnitude and distance, only considering sources with apparent G$<$20 and distance$<$1,000\,pc.
\subsection{SBI Analysis and Binary fractions}
\label{sec:big_sbi}
\begin{figure}
    \centering
    \includegraphics[width=0.9\linewidth]{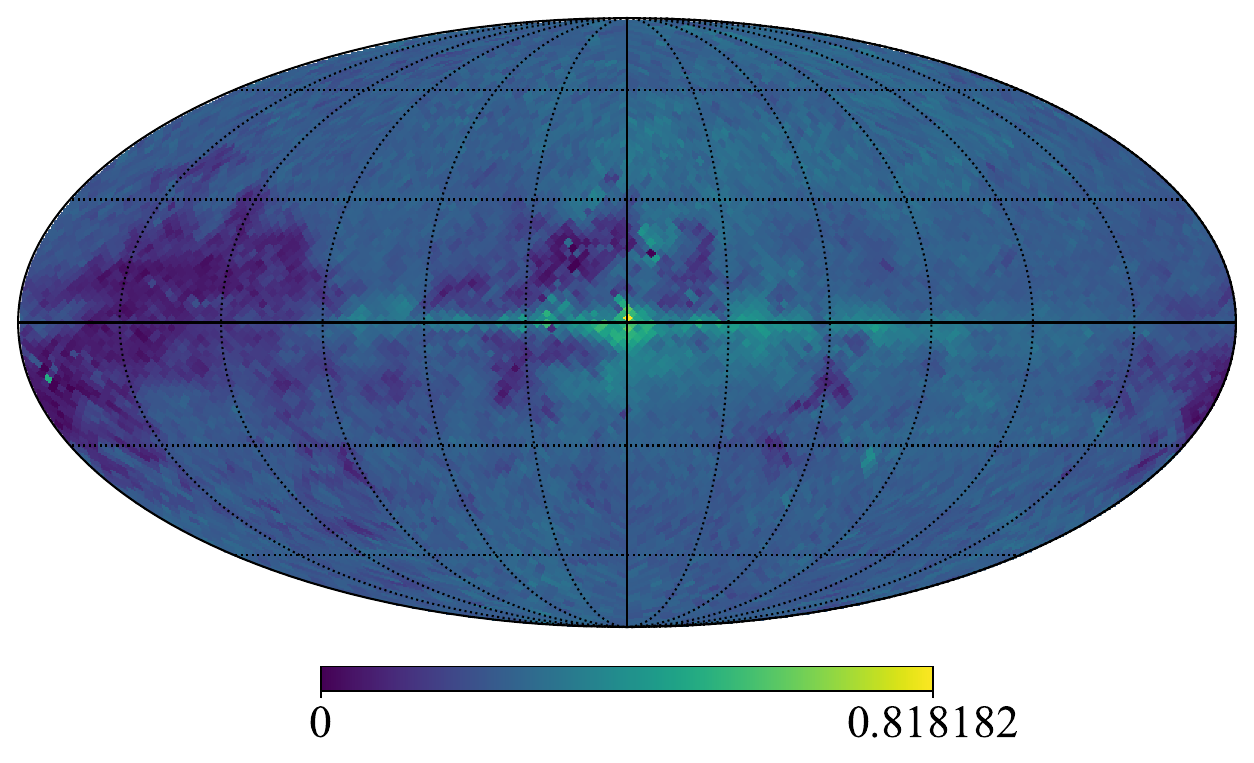}
    \caption{Colour map of binary Fraction across the entire sky.  This is presented in galactic coordinates centred at $(l,b)=(0,0)$ with $l$ increasing to the left.  There is a clear region near the galactic centre where the binary fraction is maximised and a bright point near $l=170^{\circ},b=-15^{\circ}$ which corresponds to the Taurus Molecular Cloud.}
    \label{fig:bin_fraction_sky}
\end{figure}

\begin{figure}
    \centering
     \subfigure[Histogram of recovered $q$ and {[}Fe/H{]}]{\label{fig:q_feh} \includegraphics[width=1\linewidth]{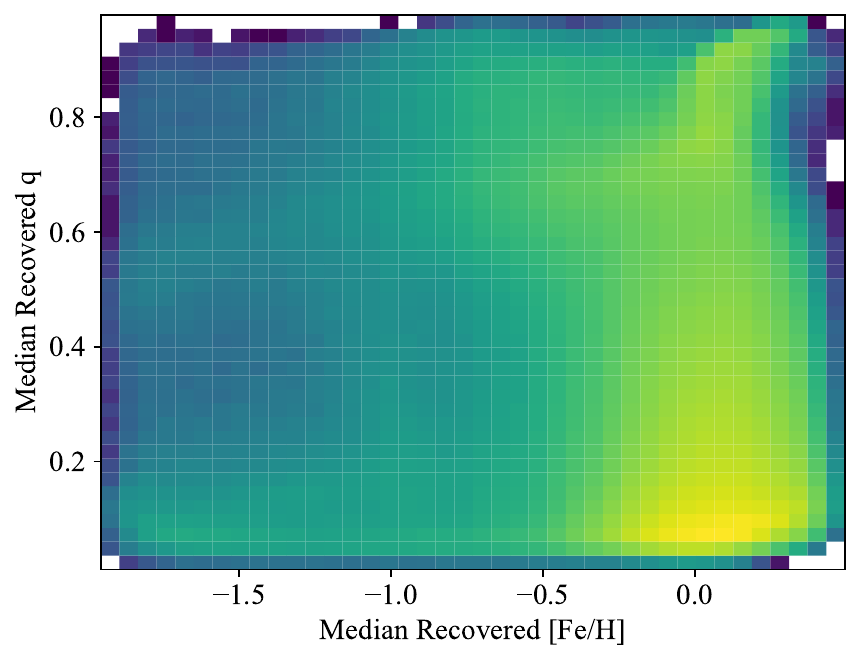}}
    \subfigure[Binary Fraction as a Function of Metallicity]{\label{fig:bf_feh} \includegraphics[width=1\linewidth]{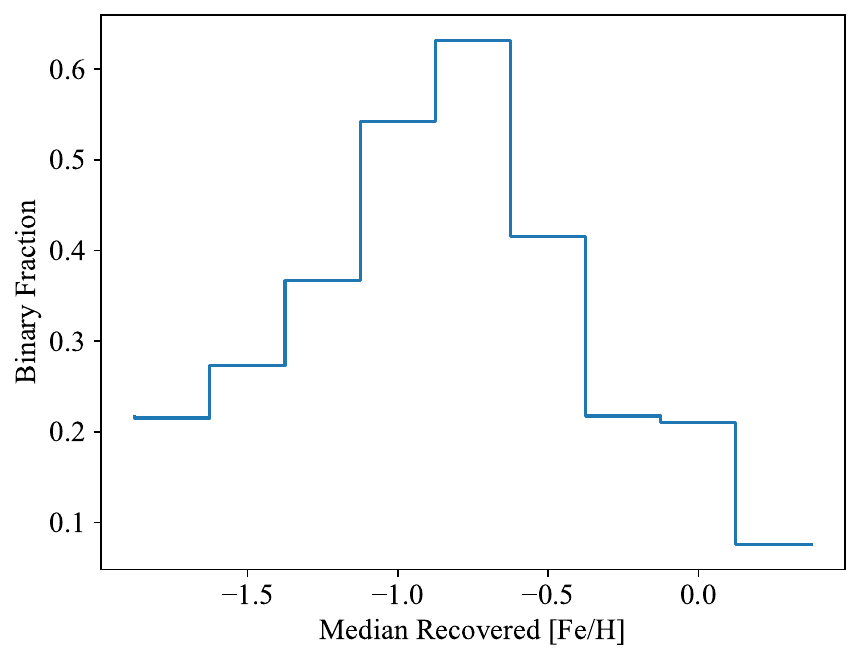}}
    \caption{Histogram of mass ratio and metallicity and binary fraction as a function of metallicity.  For higher metallicities, result is strongly anti-correlated.}
    \label{fig:bf_q_feh}
\end{figure}

The SBI method was run on Gaia sources in this region across the entire sky.  The sample of Gaia sources was distributed evenly across the sky, making it possible to investigate whether the binary fraction changes as a function of spatial position.  This is shown in galactic coordinates in Figure~\ref{fig:bin_fraction_sky}.
The binary fraction map in Figure~\ref{fig:bin_fraction_sky} shows a concentration of binaries toward the galactic centre and a lack of binaries in the galactic plane near $l=180^{\circ}$.  The perceived overabundance of binaries toward the galactic centre could be due to the $\rho$-Ophiuchus star forming region and correlates closely with the CO map from \citet{dame22}.  For sources at larger distances ($>$300\,pc), the overabundance can be explained by Gaia selection effects.  Gaia is likely to select only the brightest stars in this dense region, which are more likely to be binaries.  Outside the galactic plane, there is little variation in binary fraction. However, there is a local peak in binary fraction near galactic coordinates $l=170^{\circ},b=-15^{\circ}$ on the far left of Figure~\ref{fig:bin_fraction_sky}.  These coordinates correspond to the location of the Taurus Molecular Cloud which, along with $\rho$-Ophiuchus, is one of the closest star forming regions.  These results suggest higher binary fraction in close stellar associations.\\
The overall binary fraction is also expected to vary as a function of metallicity \citep{Moe_2019}.  Figure~\ref{fig:bf_q_feh} shows a 2D histogram comparing the recovered values of mass ratio and metallicity as well as the binary fraction as a function of metallicity.
The histogram in Figure~\ref{fig:q_feh} demonstrates the fact that most of the recovered metallicities are near solar matallicities and values below $\sim$-0.5 are rare.  The mass ratio for high metallicity values peaks at low $q$ ($\sim$0.1) and for matallicities $\sim$-0.5 it peaks at $q>0.6$.  This is reflected in the binary fraction plot shown in Figure~\ref{fig:bf_feh} which shows anti-correlation between binary fraction and metallicity which has been identified by previous studies \citep{Moe_2019}.  However, below [Fe/H]$\sim$-0.6, the binary fraction appears to increase with increasing metallicity and it is determined that binaries with [Fe/H]<-1 are quite rare.  However, as noted in Section~\ref{sec:init_test}, some of the recovered single stars at medium metallicity ([Fe/H]$\sim$-1) could actually be binary stars at lower metallicity.  These incorrectly characterized stars account for less then 5\% of the total sample but could significantly bias the result for low metallicities in which are relatively uncommon.  Overall, the binary fraction in the local stellar neighbourhood was calculated to be 23.26$\pm$0.02\%.
\subsection{Combination with Paired Catalogue}
\begin{figure}
    \centering
     \subfigure[Distributions of recovered $q$ and calculated $P$]{\label{fig:q_P} \includegraphics[width=1\linewidth]{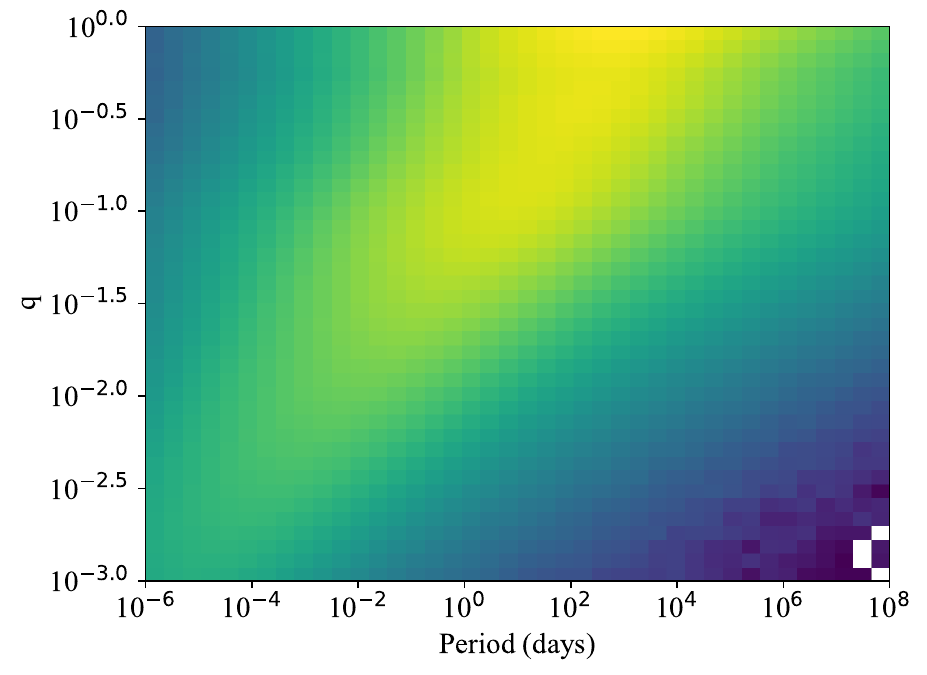}}
    \subfigure[Distribution of calculated $P$ for $q>0.2$.]{\label{fig:p_dist} \includegraphics[width=0.9\linewidth]{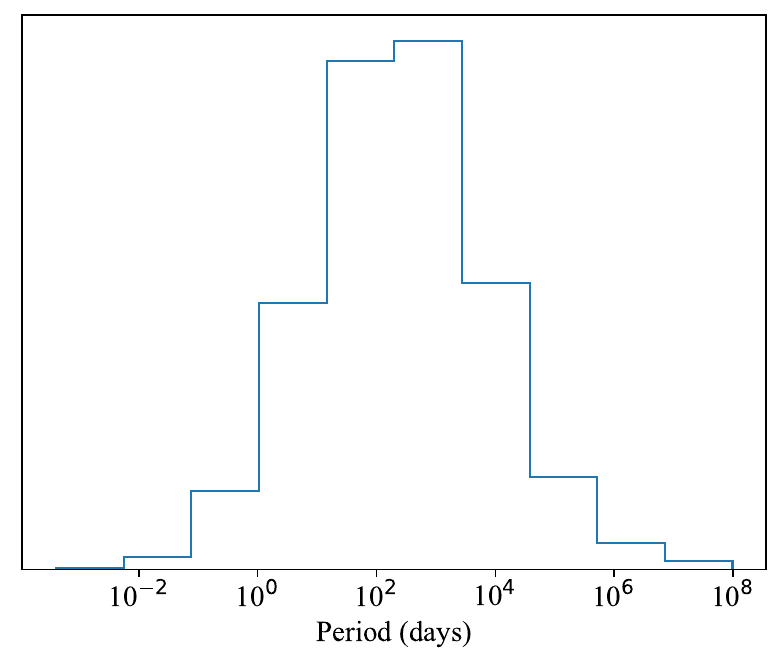}}
    \caption{Comparison of distributions of mass ratio and period distribution excluding low $q$ values.}
    \label{fig:q_P_hist}
\end{figure}

As an additional study with this data set, the sources were cross-matched with those in the \texttt{paired} catalogue with accurate radial velocity measurements \citep{quadry_chance_2022_6765903}.  The results of the SBI analysis detailed in Section~\ref{sec:big_sbi} were combined with radial velocity amplitudes from \texttt{paired} to produce a period distribution.  Assuming these systems are on circular orbits, we can make use of the relation:
\begin{equation}
    P = \frac{2\pi G}{(M_{1}+M_{2})^{2}}\left(\frac{M_{2}\rm{sin} i}{K}\right)^{3},
\label{eq:p_calc}
\end{equation}
where $M_{1}$ and $M_{2}$ are the primary and secondary masses of the system, represented as distributions from the SBI analysis where $M_{2}=qM_{1}$.  The radial velocity $K$ is sampled from a Gaussian distribution of the same size with a mean equal to the median $K$ value from \citet{quadry_chance_2022_6765903} and standard deviation taken from the 16th and 84th percentiles of $K$.  The inclination angle $i$ is unknown, so a uniform prior distribution in $\cos{i}$ is adopted.  The distributions of the recovered mass ratio and calculated periods are shown in Figure~\ref{fig:q_P}.  At low mass ratio ($q<0.2$) the calculation of the period distribution is skewed towards extremely low values (as expected by Equation~\ref{eq:p_calc}) are likely unphysical.  For $q>0.2$, the period distribution maintains a constant shape which is shown in Figure~\ref{fig:p_dist}.
The period distribution shown in Figure~\ref{fig:p_dist} peaks at periods of slightly more than 10$^{2}$ days which differs from previous studies outlined in \citet{moe17} which have the distribution peaking at over $10^{4}$ days.  This is most likely due to the sensitivity of the radial velocity survey shown in Figure\,2 of \citet{quadry_chance_2022_6765903} which demonstrates this radial velocity survey is not sensitive to companions beyond $\sim$10\,AU.
\section{Study of the Star Cluster M67}
The binary fraction of star clusters is of particular interest as this can provide information about stellar interactions and evolution.  Additionally, clusters that have been well-studied will already have constraints on the astrophysical properties of its members.  These constraints can then be applied to the SBI results to improve their accuracy.  In this section, an analysis of 367 stars in the open cluster M67 is presented.  At an age of $\sim$4\,Gyr, this is one of the oldest known open clusters.  With many Sun-like stars, it is also one of the most well-studied clusters.\\
\subsection{Forced Metallicity Distribution}
\begin{figure}
    \centering
    \includegraphics[width=1\linewidth]{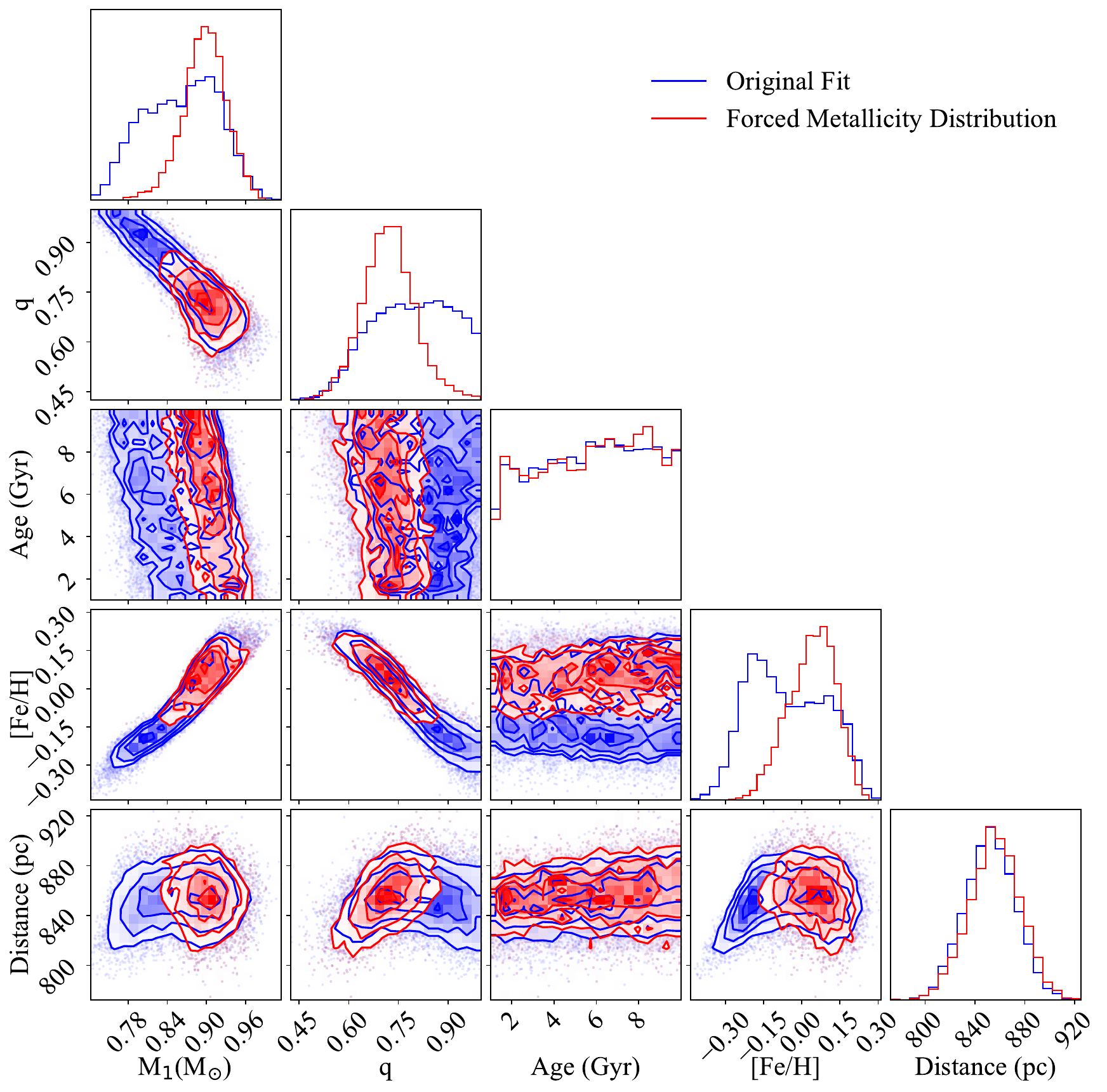}
    \caption{Comparison between parameter distributions before and after applying an assumed metallicity distribution for the cluster.  Shown for one example in M67.}
    \label{fig:new_feh}
\end{figure}

This SBI method was run on a selection of Gaia sources in M67, identified by position, proper motion and parallax measurements in the ranges specified by \citet{Ghosh_2022}.  The recovery was run as detailed in Section~\ref{sec:method} which produced a set of distributions for each star similar to that shown in Figure~\ref{fig:pairplot_ex}.  After the properties of each star have been recovered in this way, a metallicity distribution inspired by \citet{onehag2014} is applied in which a mean [Fe/H] of 0.06 and standard deviation of 0.1 are assumed.  For each star, every sample in the SBI recovery with a metallicity outside this distribution was removed.  An comparison for an example in the cluster is shown in Figure~\ref{fig:new_feh} which plots the original result from SBI and the resultant distributions when applying this forced metallicity distribution.
The new distributions shown in Figure~\ref{fig:new_feh} demonstrate a source which previously had a broad mass ratio distribution that has now been more tightly constrained an centered on slightly lower $q$ than before.  This is due to the fact that, for this particular star, the SBI recovery slightly favoured lower metallicities and higher mass ratios which would produce similar magnitudes.  By forcing the metallicity to conform to what was already known about the cluster, some of the degeneracy has been eliminated.\\
\subsection{Mass ratios and Binary Fractions}
\begin{figure}
    \centering
    \includegraphics[width=1\linewidth]{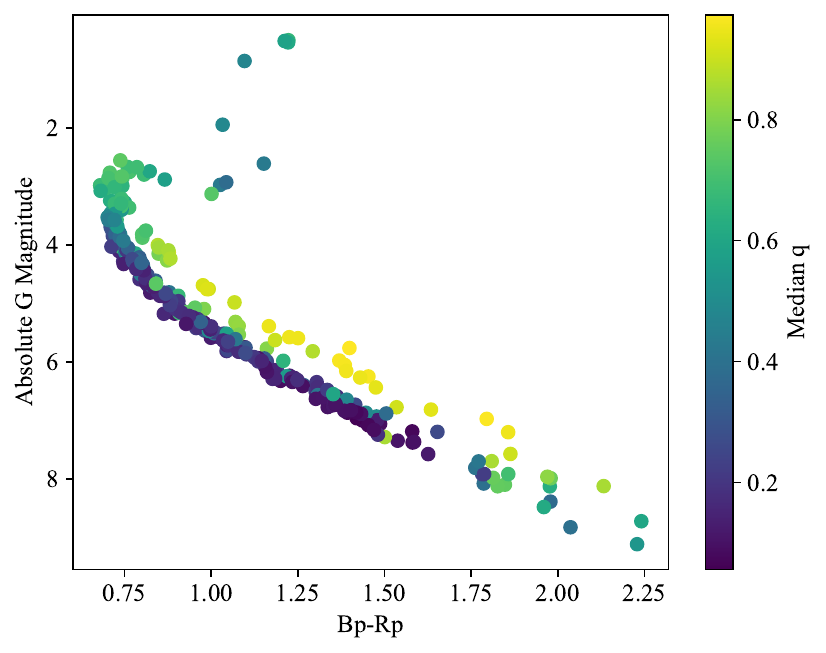}
    \caption{H-R diagram showing median mass ratio from M67.  There is a clear distinction between the single star main sequence and the set of binaries slightly above.}
    \label{fig:m67_medq}
\end{figure}

This method of a forced metallicity distribution was applied to all stars in the sample, producing more constrained estimates on stellar masses and allowing a more accurate estimation of the binary fraction.  An H-R diagram (with extinction removed) showing the median mass ratios for stars in the cluster is shown in Figure~\ref{fig:m67_medq}.  The distinction between single and binary stars is clearly evident both in the stars' positions on the H-R diagram and the recovered median masses.

\begin{figure}
    \centering
    \includegraphics[width=0.9\linewidth]{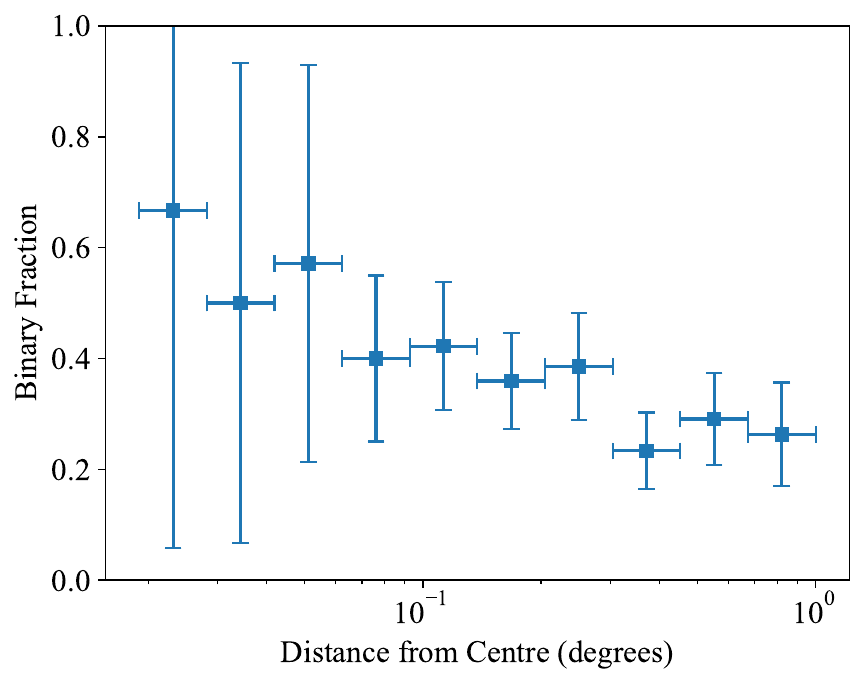}
    \caption{Binary fraction as a function of angular distance from the centre of the cluster.  The binary fraction appears to follow a downward trend indicating binaries are more common in the cluster's interior but the large uncertainties prevent any definite conclusion.}
    \label{fig:bf_sep}
\end{figure}

Using the previous definition of a binary (median $q$>0.5), the overall binary fraction of this sample is found to be 34$\pm$3\%.  This is significantly higher than the binary fraction calculated for the local stellar neighbourhood in Section~\ref{sec:big_sbi}.  The relation between this measured binary fraction and position within the cluster was also investigated.  This is represented in Figure~\ref{fig:bf_sep} which shows the binary fraction as a function of angular separation from the centre of the cluster.  The centre is defined here as (R.A.,Dec)=(132.8$^{\circ}$,+11.75$^{\circ}$.
Plotting the binary fraction as a function of separation indicates a slight downward trend which would be expected, based on previous studies indicating binaries are more common in the interior of the cluster.  However, due to the relatively low number of sources in each bin, the results have high uncertainties and the differences between binary fractions are not significant enough for us to say this trend is real.  Using the measured core radius of 8.24' \citep{Davenport_2010} and the same definition of the cluster centre, it was possible to measure the overall binary fraction inside and outside the core.  In total, 41 binaries were identified in the core out of a total sample size of 92 giving a binary fraction of 45$\pm$8\%.  Outside the core, 82 binaries were identified from a sample size of 275 giving a binary fraction of 30$\pm$4\%.  While a higher binary fraction was found towards the centre of the cluster, the large uncertainties on both binary fractions again prevent a definite conclusion.  It is possible the higher binary fraction near the centre is due to a Gaia selection effect.
\section{Summary and Conclusions}
This work has shown that it is possible to recover parameters of main sequence stars through photometry alone.  The primary mass and metallicity can be recovered to a high degree of accuracy while the age has proven difficult to constrain.  The binary mass ratio can also be recovered accurately in most cases and, when applied to real data, can be used to search for binaries.\\
Using an assumed definition of a binary, it is then possible to estimate the binary fraction of a sample.  Using a simulated data set, regions in colour/magnitude space were identified in which this method produces more accurate results.  This is possibly due to the higher density of the training set in these regions.\\
Focusing in on the region on the H-R diagram where the method produces accurate results, it was possible to analyse the binary fractions of the local stellar neighbourhood as functions of position on the sky and metallicity.  Overall, the binary fraction was found to change little with position in the sky but this study has reproduced the expected anti-correlation between binary fraction and metallicity.  It has been shown that these results can be combined with radial velocity studies to produce period distributions with a shape that agrees with previous studies but peaks at lower periods than expected.  However, this can be attributed to lack of sensitivity at wide separations.\\
Through the study of M67, it has been shown that applying a known distribution of a star's metallicity can greatly improve the precision of the recovery of other parameters including mass ratio.  A binary fraction was calculated which is almost constant regardless of distance from the centre but exhibits a slight downward trend.  Out binary fraction for M67 was calculated to be 34$\pm$3\% which is higher than the 23.26$\pm$0.02\% calculated for field stars.  This suggested a similar result to previous studies \citep{offner22} which found that binaries are more common in environments with higher stellar density.\\
Moving forward, this project enables a more complete picture of the binary fraction of sources in the Gaia data.  Radial velocity and astrometric surveys are primarily concerned with detecting dark companions and, with this extra knowledge of unreolved bright companions, these surveys will be able to place better constraints on hidden companions.

\section*{Data Availability}
The code used to recover astrophysical parameters with SBI is available in a github repository (\href{https://github.com/awallace142857/sbi_code}{https://github.com/awallace142857/sbi\_code}) which also includes some Gaia data.  Due to the size of the data files, the remaining data are available from the corresponding author on reasonable request.

\section*{Acknowledgements}
A.L.W. acknowledges the contributions of Andrew Casey who helped start this project and edit this paper in its early stages.  A.L.W. also thanks Anthony Brown, Daniel Foreman-Mackey and Quadry Chance for their helpful suggestions regarding the discussion and overall science goals of the study.  This work was funded by the Australian Research Council Discovery Grant DP210100018.




\renewcommand\refname{References}
\bibliographystyle{mnras}
\bibliography{references} 








\bsp	
\label{lastpage}
\end{document}